\documentclass[letterpaper,twocolumn,preprintnumbers,secnumarabic,amsmath,amssymb,superscriptaddress,nofootinbib]{revtex4-1}
\usepackage{amssymb,amsmath}
\usepackage{ulem} %for strikethrough
\usepackage{epsfig}
\usepackage{bm} %math bold symbols
\usepackage{booktabs} %some alignment stuff
\usepackage{array}
\usepackage{latexsym}
\usepackage{mathrsfs}
\usepackage{graphicx}
\usepackage{color}
\usepackage{datetime}

\usepackage{lettrine}

\usepackage{tablefootnote}
\usepackage{multirow}

\usepackage{verbatim}
\usepackage{chngpage} % allows for temporary adjustment of side margins

\usepackage{psfrag}

\usepackage{mciteplus}

\usepackage[colorlinks=true,      linkcolor=blue,      urlcolor=blue,      
            filecolor=blue,      citecolor=blue,       pdfstartview=FitH,     
						pdfpagemode=UseNone,      bookmarksopen=true]{hyperref}  % LINKS
\usepackage[all]{hypcap}     %should be loaded AFTER hyperref, which should otherwise be loaded last.

\def\({\left(}
\def\){\right)}
\def\[{\left[}
\def\]{\right]}

\def\barray{\begin{array}}
\def\earray{\end{array}}
\def\be{\begin{equation}}
\def\ee{\end{equation}}
\def\bea{\begin{eqnarray}}
\def\eea{\end{eqnarray}}
\def\bal{\begin{align}}
\def\eal{\end{align}}

\def\-{\,-\,}
\def\={\,=\,}
\def\+{\,+\,}

\definecolor{cardinal}{rgb}{0.6,0,0}
\definecolor{darkgreen}{rgb}{0,0.4,0}
\definecolor{golden}{rgb}{0.92, 0.7, 0}
\definecolor{midnight}{rgb}{0, 0, 0.5}
\definecolor{darkblue}{rgb}{0, 0, 0.7}
\definecolor{purple}{rgb}{0.5, 0, 0.5}

\def\cB{{\cal B}}

\def\cM{{\cal M}}

\def\cO{{\cal O}}

\def\tr{\mathop{\mathrm{tr}}\nolimits}

\usepackage{graphicx}

\usepackage{tikz}
\usetikzlibrary{decorations.markings}
\tikzset{->-/.style={decoration={
			markings,
			mark=at position #1 with {\arrow{stealth}}},postaction={decorate}}}
\usepackage{adjustbox}
\usepackage{pgfplots}
\pgfplotsset{compat=1.11}
\usepgfplotslibrary{fillbetween}
\usetikzlibrary{intersections}

\pgfdeclarelayer{bg}
\pgfsetlayers{bg,main}

\begin{document}
%%%%%%%%%%%%%%%%%%%%%%%%%%%%%%
\title{An Alliance in the Tripartite Conflict over Moduli Space}

\author{Yixuan Li}
\affiliation{Institut de Physique Th\'eorique, Universit\'e Paris-Saclay, CNRS, CEA, Orme des Merisiers 91191, Gif-sur-Yvette CEDEX, France\\ \vspace*{1mm}{\rm \textsf{ yixuan.li@ipht.fr} }}

\begin{abstract}
We investigate three proposals of distance on the moduli space of metrics: (1) a distance derived from the symplectic form of phase space, (2) a distance obtained by moving BPS objects at small velocity, (3a) a distance proposed by DeWitt and (3b) the distance used in the context of the generalised Swampland distance conjecture.
%, one of which being used in the context of the generalised Swampland distance conjecture.
%
In particular, we calculate these distances on a space of geometries that have the same asymptotics as the supersymmetric black hole in five dimensions. 
These moduli spaces contain a locus where there exists an infinite tower of massless particles, which emerges at \textit{finite} distance according to proposals (1) and (3a), and at \textit{infinite} distance from proposal (2) and (3b):
distances (1) and (3) agree, and they disagree with distance (3b).

%Using the third notion, we compute the distance in moduli space from a geometry in the bulk moduli space to a solution whose metric approaches that of the black hole as seen from the outside.
%
%We find that this distance is finite, and the result matches a previous computation by the same author using the first notion of distance.
%
%Thus, our computation gives an example of an infinite tower of massless states which emerges at finite distance in moduli space.

%Compared to the second notion of distance however, these results seem to be in tension with a previous paper in the literature..
\end{abstract}
%\vspace*{-111mm}

\maketitle

%%%%%%%%%%%%%%%%%%%%%%%%%%%%%%%%%%%%%%%%%%%%%%%%%%%%%%%%%%%%%%%%%%%%%%%%%%%%%%%%%%%%%%%%%%%%%%%%%%
%%%%%%%%%%%%%%%%%%%%%%%%%%%%%%%%%%%%%%%%%%%%%%%%%%%%%%%%%%%%%%%%%%%%%%%%%%%%%%%%%%%%%%%%%%%%%%%%%%
%%%% 	Introduction
%%%%%%%%%%%%%%%%%%%%%%%%%%%%%%%%%%%%%%%%%%%%%%%%%%%%%%%%%%%%%%%%%%%%%%%%%%%%%%%%%%%%%%%%%%%%%%%%%%
%%%%%%%%%%%%%%%%%%%%%%%%%%%%%%%%%%%%%%%%%%%%%%%%%%%%%%%%%%%%%%%%%%%%%%%%%%%%%%%%%%%%%%%%%%%%%%%%%%
\section{Introduction}

%\paragraph{Different notions of moduli space, but from the same UV-complete theory.}
%The moduli space of flux vacua is a space of EFTs, whereas the moduli space of black hole microstates is a space of solutions within the same EFT. However, from the string theory perspective, both moduli spaces are subsectors of a solution space of the same UV-complete theory.

The moduli space of solutions of a given theory describes a set of solutions parameterized by continuous parameters.
For our intuition, it is practical to define a distance on the moduli space so that similar solutions are close one to another in the space of solutions.
There are, in the literature, three ways to define distances on moduli spaces which come from a priori different formulas and viewpoints.

\textbf{1. The phase-space distance.}
%The moduli space of microstate geometries is thought as a classical phase space upon which \textit{geometric quantization} gives the number of quantum microstates of the black hole.
A first way to think about moduli space is to start with a semi-classical theory; one can define a classical phase space upon which geometric quantization \cite{Ritter:2002zg,Echeverria-Enriquez:1998umj,Berman:2022acl} gives the number of quantum states in a given region of the phase space. For instance, this is the point of view one can take in order to count the number of quantum black-hole microstates that a given set of supergravity solutions (parameterized by continuous parameters) account for \cite{Maoz:2005nk,Rychkov:2005ji,deBoer:2008zn,Mayerson:2020acj}. In detail, given a Lagrangian theory characterizing a set of fields, $\phi^A$, one defines the symplectic form of the theory, $\Omega$, from the Crnkovi\'c-Witten-Zuckerman formalism \cite{Witten:1986qs,Crnkovic:1987tz}:
\be
\Omega = \int d\Sigma_l \,\, \delta\left( \frac{\partial L}{\partial (\partial_l \phi^A)}\right) \wedge \delta \phi^A \, ,
\ee
where the integral is performed over a Cauchy surface $\Sigma$ in space-time.
The fundamental quantity in phase space is given by the symplectic form, $\Omega$, and there is not necessarily a notion of distance on it. Nevertheless, the symplectic structure of the manifold is sometimes compatible with an almost complex structure, $J$; then one can define a K\"ahler metric out of $\Omega$ and $J$:
\begin{align} %\label{eq:phase_metric}
    G_{\mathrm{phase}}=\Omega(\cdot,J\cdot) \,.
\end{align}
Given a path, $\gamma$, linking two points $p_1$ and $p_2$ on moduli space, one can define the distance along the path, $\gamma$, between the two solutions:
\begin{align} \label{eq:phase_distance}
\Delta_\mathrm{phase}(p_1,p_2)=\int_{\gamma}\sqrt{G_{ab}\frac{\partial x^a}{\partial \tau}\frac{\partial x^b}{\partial \tau}} \mathrm{d}\tau \,.
\end{align}

\textbf{2. The low-velocity distance.}
%Another point of view is the Michelson-Strominger distance, which corresponds to adiabatic transformations of the metric. \YL{Add formula and comments.}
In the context of a solution describing multiple static, extremal Reissner-Nordstr\"om black holes, one can define a distance on the moduli space parameterized by the distance between the black holes \cite{Ferrell:1987gf,MANTON198254}. Because each black hole sources electric and gravitational forces that cancel out one another, the overall system is stable, and any configuration of those black holes described by their coordinate in space, $\Vec{x_j}$, solves the equations of motion. 
One can allow the black holes to move with small velocities $\Vec{v_j}\equiv \frac{\mathrm{d}\Vec{x_j}}{\mathrm{d}t}$, and find the expansion of the action up to order $\mathcal{O}(v^2)$. Then, one rewrites this effective action so that it involves the velocities, $\Vec{v_i}$, as an overall factor of the Lagrangian \cite{Michelson:1999dx,Douglas:1997vu,Kaplan:1997gk,Michelson:1997vq}:
\begin{align}
S_{\mathrm{eff},\, \cO(v^2)}=\int dt (v_i)^a (v_j)^b (G^{ij})_{ab} \,,
\end{align}
from which we deduce the metric of the $(d \,n)$-dimensional (where $d$ is the dimension of the physical space and $n$ is the number of black holes) configuration space of the positions $\Vec{x_j}$:
\begin{align} \label{eq:low-velocity_distance}
ds^2_{\mathrm{low-v.}}= (G^{ij})_{ab} \, \mathrm{d}{x_i}^a \, \mathrm{d}{x_j}^b \,.
\end{align}
Then the moduli space is a submanifold embedded in this configuration space, as it satisfies additional constraints, such as the conservation of the angular momentum of the whole system of black holes.

This notion of distance can be extended to other configurations of supersymmetric objects that satisfy a no-force condition. A geodesic in this moduli space represents the trajectories of the individual components of the dynamical system in physical space-time \cite{Ferrell:1987gf}.
%So one can extend this metric to some microstates of the supersymmetric black hole.
%\YL{geodesic}

\textbf{3. The DeWitt distances.} %, or `generalized' distance
Another possibility to define a metric on the moduli space of metrics has been formulated by DeWitt \cite{DeWitt:1967yk}. The original DeWitt distance applies to the moduli space of induced metrics on Cauchy slices, but one can generalise it to moduli space of (Riemannian) metrics on the entire spacetime manifold. There are two possible generalisations:\\
\textbf{3a.} \emph{The DeWitt distance without the volume factor.} 
Given a metric $g_{\mu\nu}$ of a spacetime $\cM$, the distance on a path, $\gamma$, parameterized by $\tau$ and on which the metric variations are transverse-traceless, can be defined as \cite{gilmedrano1992riemannian}
\begin{align} \label{eq:DeWitt_distance_without_volume}
\Delta_\textrm{DeWitt 1}=c \int_{\tau_{i}}^{\tau_{f}}\left( \int_{\cM} \sqrt{g} \operatorname{tr}\left[\left(g^{-1} \frac{\partial g}{\partial \tau}\right)^{2}\right]\right)^{\frac{1}{2}} \mathrm{d} \tau \,,
\end{align}
where $c$ is a constant of order 1, depending on the dimension of $\cM$.\\
\textbf{3b.} \emph{The DeWitt distance with the volume factor.}
Let $V_{\cM}=\int_{\cM} \sqrt{g}$ be the volume of $\cM$. Another distance on the same path $\gamma$ can be defined as
\begin{align} \label{eq:generalized_distance}
\Delta_\textrm{DeWitt 2}=c \int_{\tau_{i}}^{\tau_{f}}\left(\frac{1}{V_{\cM}} \int_{\cM} \sqrt{g} \operatorname{tr}\left[\left(g^{-1} \frac{\partial g}{\partial \tau}\right)^{2}\right]\right)^{\frac{1}{2}} \mathrm{d} \tau \,,
\end{align}
where $c$ is a constant of order 1, depending on the dimension of $\cM$.
This distance has been used in the context of the Swampland programme, in order to formulate the Generalized Distance Conjecture \cite{Lust:2019zwm}.
The distance (\ref{eq:generalized_distance}) boils down to the moduli space distance of the scalar fields in the case of Calabi-Yau compactifications on 4-dimensional Minkowski space \cite{Candelas:1990pi}, thus making the link with the original Swampland distance conjecture \cite{Ooguri:2006in}.

The ``Swampland'' perspective on moduli space is that given a set of scalar fields of a $d$-dimensional effective field theory, $\phi^i$, their kinetic terms, $G_{ij}$, that appear in the action written in the $d$-dimensional Einstein frame
\begin{align}
S = \int \mathrm{d}^d x \,\sqrt{-g} \left[\frac{R}{2} - G_{ij}\left( \phi^i \right) \partial \phi^i \partial \phi^j + ... \right] \,,
\end{align}
defines a metric on moduli space of the $\phi^i$'s. The geodesic distance on moduli space, $\Delta$, is then used to delimit the domain of validity of the space of effective field theories (EFT's) \cite{Klaewer:2016kiy,Baume:2016psm}. Indeed, as one moves away from a point, $p_0$, in the bulk moduli space, there should exist an infinite tower of states with an associated mass scale, $M(p)$, such that 
\begin{align}
\label{eq:Swampland_distance_conjecture}
M(p) \sim M(p_0) \, e^{-\alpha \Delta(p_0,p)} \,,
\end{align}
with $\alpha$ a constant of order 1 in Planck units. Thus, the effective field theory defined by $p$ breaks down if one moves a few Planck units away from the original effective theory, $p_0$.
In \cite{Lust:2019zwm}, the fact that the kinetic terms define a metric on moduli space was extended to the kinetic terms and the moduli space of all dynamical fields, and was linked to the DeWitt distances (\ref{eq:DeWitt_distance_without_volume},\ref{eq:generalized_distance}).
\newline

The previous three/four definitions come from different viewpoints, and are used in different contexts.
Indeed, the moduli space of flux vacua is a space of effective field theories, whereas the moduli space of multiple extremal black holes or of black hole microstates is a space of solutions within the same EFT. However, from the string theory perspective, all these moduli spaces are subsectors of a solution space of the same UV-complete theory. An obvious question is whether these three different notions of moduli space measure the same distance.

In this letter, we compare these three different notions on a specific space of metrics: the so-called five-dimensional \textit{bubbling geometries}, or equivalently, the four-dimensional \textit{multi-centered solutions}. 
%Spoiler alert: as suggested by the title,

\emph{Bubbling geometries} \cite{Bena:2005va,Bena:2006is,Warner:2019jll} are smooth, supersymmetric solutions of five-dimensional, ungauged, $\mathcal{N}=2$ supergravity, coupled to vector multiplets.
They have the same asymptotics and asymptotic charges as the three-charge, five-dimensional, supersymmetric black hole (the BMPV black hole \cite{Breckenridge:1996is}). %They are smooth in five dimensions, and can have $\mathbb{R}^{1,3}\times S^1$ or $\mathbb{R}^{1,4}$ asymptotics. 
Unlike the extremal black hole whose horizon lies at the bottom of an infinitely-long $\mathrm{AdS}_2$ throat, the bubbling solutions have a smooth cap at the bottom of a long, but finite $\mathrm{AdS}_2$ throat.
As such, they are sometimes considered as coherent superpositions of the black hole microstates \cite{Bena:2006kb,Bena:2007kg}.
Much like their corresponding black hole, the bubbling geometries can also be constructed with $\mathbb{R}^{1,3}\times S^1$ asymptotics; reduced to four dimensions, they correspond to a class of \emph{multi-centered solutions} \cite{Bates:2003vx,Behrndt:1997ny,Denef:2000nb}, which still have the same asymptotics and asymptotic charges as the four-dimensional black hole, but are singular from a 4D perspective.

One can think the multi-centered solutions as coming from splitting the black hole's asymptotic charges $(Q_1,Q_2,Q_3)$ into local charges at different locations in space, denoted \textit{centres}. The centres are located at the coordinates $\lambda \, \Vec{d_j}$ in $\mathbb{R}^3$, with $j$ enumerating the centres and $\lambda$ a positive parameter. If one wishes the metric to be smooth in five dimensions, then the relative positions of the centres are constrained.

From the classical point of view, there is no obstruction to send the parameter $\lambda$ to $0$, so that the coordinates of the centres approach the origin of $\mathbb{R}^3$ while approximately keeping their collective shape \cite{Bena:2006kb}. 
This limit in the moduli space of bubbling geometries is called the \emph{scaling limit}.
In this limit, in gravity, the length of the bubbling geometry's throat increases towards infinity, while the geometry of the cap remains fixed.
This limit constitutes a well-defined path in moduli space. So, given a solution in the bulk moduli space, $p(\lambda_0)$, it is possible to compute its distance to a solution approaching the scaling limit, $p(\lambda)$, with $\lambda \rightarrow 0$. 
In this letter, we compare the distance to the scaling limit from $p(\lambda_0)$ according to the three definitions of distance on moduli space mentioned above.
In particular, the new computation we do in this letter is the one using the formula with the DeWitt distances (\ref{eq:DeWitt_distance_without_volume}) and (\ref{eq:generalized_distance}).

Before computing the distance to the scaling limit in the moduli space of bubbling geometries, let us recall a good illustrative example where the second DeWitt distance (\ref{eq:generalized_distance}) is used in the context of the Swampland: Take a family of $\textrm{AdS}_p\times\textrm{S}^q$, where the radius of AdS space fixes the size of the sphere \cite{Lust:2019zwm}. The limit in moduli space where the value of the cosmological constant vanishes lies at infinite distance from the bulk moduli space. Besides, while going to that limit, an infinite tower of Kaluza-Klein (KK) modes of the sphere becomes exponentially massless, in accordance with the Swampland distance conjecture.

Now, take a family of warped geometries that are asymptotically $\textrm{AdS}_p\times X$, where $X$ is a compact manifold. To compute the exact mass of the KK modes of $X$, one needs to take a scalar deformation of the metric, solve the wave equations, and the quantized energies measured at spatial infinity give the masses of the KK tower \cite{Bena:2018bbd}. %And if one is satisfied with approximate masses of the KK modes, one can assess them through the redshift factor.

For our bubbling geometries with a long $\mathrm{AdS}_2 \times \mathrm{S}^3$ throat, any energy excitation at the bottom of the throat is redshifted when one measures it at spatial infinity. In the scaling limit, the redshift becomes stronger and stronger, so the energy measured at spatial infinity gets more and more suppressed. Moreover, this decay is exponential with respect to the throat length, $L_{\mathrm{throat}}$ \cite{Li:2021gbg}:
\begin{align}
%\label{eq:masstower_Lthroat}
M(L_{\mathrm{throat}}) \underset{\lambda \rightarrow 0}{\sim}
 \exp\( -\frac{L_{\mathrm{throat}}}{(Q_1Q_2Q_3)^{1/6}} \) \,.
\end{align}
Knowing the Swampland distance conjecture (\ref{eq:Swampland_distance_conjecture}), it appeared natural to propose that the distance to the scaling limit in moduli space is given by the argument of the decreasing exponential, up to some constant factor of order 1 \cite{Li:2021gbg}:
\begin{align}
\label{eq:exponential_distance}
\Delta_{\textrm{exponential}}(\lambda_0,\lambda)= \frac{|L_{\mathrm{throat}}(\lambda)-L_{\mathrm{throat}}(\lambda_0)|}{(Q_1Q_2Q_3)^{1/6}} \,.
\end{align}
Thus, with $\Delta_{\textrm{exponential}}$, the distance to the scaling limit would be infinite, as the throat length increases to infinity to match that of the extremal black hole.

However, this notion of moduli space distance can only be used if one could show that every tower of massless states emerges at infinite distance: namely, that the converse of the Swampland distance conjecture applies in this limit. 

But it was also shown from \cite{Li:2021gbg} that the distance to the scaling limit according to the phase-space definition (\ref{eq:phase_distance}) would give a \textit{finite} distance. This distance was derived at weak string coupling from quiver quantum mechanics, and, according to \cite{deBoer:2008zn}, can be extrapolated to strong coupling thanks to a non-renormalization theorem \cite{Denef:2002ru,deBoer:2008zn}. In the scaling limit, the relevant metric component on moduli space, $G_{\lambda \lambda}$, blows up like $1/\lambda$, so the phase-space distance behaves like
\begin{align} \label{eq:Delta_phase_scaling}
\Delta_{\mathrm{phase}}(\lambda_0,\lambda) = \int_{\lambda}^{\lambda_0} d\lambda' \, \sqrt{G_{\lambda \lambda}}
\propto \sqrt{\lambda_0}-\sqrt{\lambda} \approx \sqrt{\lambda_0} \,
\end{align}
in the vicinity of the scaling limit.

The phase-space distance indicates that an infinite tower of massless modes can emerge at finite distance in moduli space. 
However, the phase-space distance is not the one used to formulate the Generalized Distance Conjecture \cite{Lust:2019zwm}.
Therefore, in this letter, we compute the distance to the scaling limit according to the DeWitt distances. 

In Section \ref{sec:multicentered_solutions}, we review some properties of bubbling geometries. In Section \ref{sec:computation}, we show that the first DeWitt distance (without the volume factor) for the scaling limit matches the phase-space distance (\ref{eq:Delta_phase_scaling}), while the second DeWitt distance (with the volume factor) matches the ``exponential'' distance (\ref{eq:exponential_distance}).
%Therefore, the computation provides a counter-example to the converse of the Swampland distance conjecture, generalised for metrics. 
In Section \ref{sec:discussion}, we confront these two distances with a distance computed in the literature \cite{Ferrell:1987gf,Michelson:1997vq,Maldacena:1998uz} using the low-velocity distance.

\section{Bubbling geometries}
\label{sec:multicentered_solutions}

The metric of the bubbling geometries is of the form
\begin{align}
ds_{5}^2 = &-\left(Z_M\right)^{-2} \left(dt+\omega \right)^2 +  \frac{Z_M}{V} \left(d\psi + A\right)^2 \nonumber\\
&+ V Z_M \,\biggl[\,d\rho^2 + \rho^2 \left( d\theta^2 + \sin^2 \theta \, d\phi^2 \right)\,\biggr]\,,
\label{eq:BubblingMetric}
\end{align}
where the warp factor $Z_M\equiv\( Z_1Z_2Z_3 \)^{1/3}$ is the geometric mean of the functions $(Z_1,Z_2,Z_3)$ which encode the three asymptotic charges of the black hole.

While the black hole has warp factors $Z_I$ sourced by a charge $Q_I$ at the origin of the four-dimensional space and which is of the form
\begin{align} \label{eq:harmonic_func_general}
    Z_I=1+\frac{Q_I}{\rho} \,,
\end{align}
the multi-centered solution are determined by eight harmonic functions on the base $\mathbb{R}^3$, $(V,K^I;L_I,M)\equiv\mathbf{H}$, that depend on the location of their poles, in the following generic form:
\begin{align}
H=h_\infty+\sum_{j=1}^{n}\frac{h_j}{|\Vec{\rho}-\lambda \, \Vec{d_j}|} \,.
\end{align}
The integer $n$ denotes the number of centres. The coefficient $h_\infty$ is the asymptotic value of the harmonic function $H$; collectively, the $h_\infty$ are chosen to be $\mathbf{h}_\infty \equiv (v_\infty,l^I_\infty;k^I_\infty,m_\infty)=(1,1,1,1;0,0,0,m_\infty)$ for $\mathbb{R}^{1,3}\times S^1$ asymptotics. 
The coefficient $h_j$ is the charge associated to the centre $j$; collectively, the $\mathbf{h}_j\equiv (v_j,l^I_j;k^I_j,m_j)$ satisfy 
\begin{align} \label{eq:smoothness_cond_GH}
l_j^I \=-\frac{|\varepsilon_{IJK}|}{2} \frac{k_j^J k_j^K}{q_j}\, ,\qquad m_j \= \frac{|\varepsilon_{IJK}|}{12} \frac{k_j^I k_j^J k_j^K}{q_j^2}\, ,
\end{align}
with $\varepsilon_{IJK}$ being the Levi-Civita symbol.
The equations (\ref{eq:smoothness_cond_GH}) guarantee the smoothness of the bubbling geometries in five dimensions.
Besides, the absence of Dirac-Misner strings (which give of closed time-like curves) in bubbling geometries leads to constraints on the relative positions of the centres, $\rho_{ij}\equiv | \lambda \, \Vec{d_j} - \lambda \, \Vec{d_i} |$, the so-called \textit{bubble equations}, or \textit{Denef equations} \cite{Denef:2000nb,Bena:2006is}:
\begin{align}
\label{eq:BubbleEquations}
\sum_{j=1}^n \frac{ \langle \mathbf{h}_i , \mathbf{h}_j \rangle}{\rho_{ij}} \= \langle\mathbf{h}_\infty , \mathbf{h}_i \rangle \, , \qquad \textrm{for  } i\=1,\ldots n\,.
\end{align}
Here, the symplectic product, $\langle\cdot,\cdot\rangle$, for $\mathbf{A}=(A^0,  A^I; A_I, A_0)$ and $\mathbf{B}=(B^0,  B^I; B_I, B_0)$ is defined as $\langle \mathbf{A},\mathbf{B} \rangle \equiv A^0 B_0-A_0 B^0 + A^I B_I-A_I B^I$.

The metric of the bubbling geometries (\ref{eq:BubblingMetric}) comprises warp factors (and angular momentum) whose building blocks are the harmonic functions $(V,K^I;L_I,M)$:
\begin{align}
Z_I \= L_I \+ \frac{|\varepsilon_{IJK}|}{2} \frac{K^J K^K}{V} \,.
\label{eq:Z_I_intermsof_harmonic}
\end{align}
Asymptotically, the different local charges $\mathbf{h}_j\equiv (v_j,l^I_j;k^I_j,m_j)$ put on the centres develop into the black hole's asymptotic charges, $(Q_1,Q_2,Q_3)$, in the following fashion:
\begin{align}
\label{eq:Asymptotic_charges_5d}
Q_I \= \sum_{j=1}^n l_j^I \+ |\varepsilon_{IJK}|\, \sum_{(i,j)=1}^n k_i^J k_j^K \, .
\end{align}

%The centres are located at the coordinate $\lambda\, \Vec{d_j}$, with $j$ enumerating the centres. In the 3-dimensional base space, the distance between the point $\Vec{\rho}$ and the $j^{\mathrm{th}}$ centre is denoted $\rho_j\equiv |\Vec{\rho}-\lambda \, \Vec{d_j}|$.

%%%%%%%%%%%%%%%%%%%%%%%%%%%%%%%%%%%%%%%%%%%%%%%%%%%%%%%%%%%%%%%%%%%%%%%%%%%%%%%%%%%%%%%%%%%%%%%%%%
%%%%%%%%%%%%%%%%%%%%%%%%%%%%%%%%%%%%%%%%%%%%%%%%%%%%%%%%%%%%%%%%%%%%%%%%%%%%%%%%%%%%%%%%%%%%%%%%%%
%%%% 	The computation
%%%%%%%%%%%%%%%%%%%%%%%%%%%%%%%%%%%%%%%%%%%%%%%%%%%%%%%%%%%%%%%%%%%%%%%%%%%%%%%%%%%%%%%%%%%%%%%%%%
%%%%%%%%%%%%%%%%%%%%%%%%%%%%%%%%%%%%%%%%%%%%%%%%%%%%%%%%%%%%%%%%%%%%%%%%%%%%%%%%%%%%%%%%%%%%%%%%%%
\section{The DeWitt distances between deep-throat geometries}
\label{sec:computation}

%To compute the geometric distance formula,

\textit{a. The metric and its inverse in the throat region.}
As the cap and the asymptotics are fixed in the scaling limit, we only need to compare the distance between solutions with throats of different lengths, using the DeWitt distance (\ref{eq:DeWitt_distance_without_volume}).%
\footnote{The geometry in the cap and in the asymptotic flat region does not depend on the scaling parameter, $\lambda$, so the $\frac{\partial g}{\partial \lambda}$ term in (\ref{eq:DeWitt_distance_without_volume}) vanishes in these regions.}

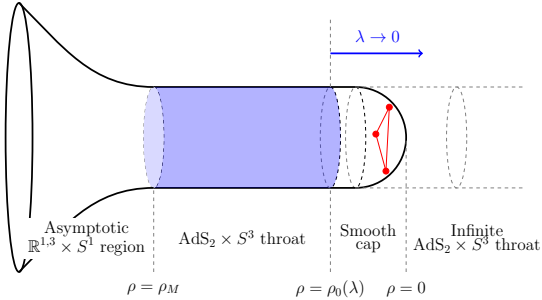
\begin{figure}[h]
	\label{fig:deep-throat_geometry}
	\begin{adjustbox}{max totalsize={0.4\textwidth}{\textheight},center}
		\begin{tikzpicture}
			\begin{scope}[shift={(0,0)}]
			    % Cap
				\draw[black,line width=1.5] (0, 1.5) arc (90:-90:1.5);
				%\draw [gray, thick,dashed](0.0,0) circle [x radius=0.3, y radius=1.5];
				\draw[black, thick,dashed] (0,1.5) arc (90:-90: 0.3 and 1.5 );
				\draw[gray, thick,dashed] (0,-1.5) arc (-90:-270: 0.3 and 1.5 );
				%
				% Throat
				\draw [ black, line width=1.5] (0, 1.5) --(-6,1.5);
				\draw [ black, line width=1.5] (0, -1.5) --(-6,-1.5);
				% Infinite throat
				\draw [gray, thick, dashed] (0,1.5) -- (5, 1.5);
				\draw [gray, thick, dashed] (0,-1.5) -- (5, -1.5);
				\draw [gray, thick,dashed, ] (3,0) circle [x radius=0.3, y radius=1.5];
				%\draw [gray, thick,dashed, ] (-0.75,0) circle [x radius=0.3, y radius=1.5];
				\draw[black, thick,dashed] (-0.75,1.5) arc (90:-90: 0.3 and 1.5 );
				\draw[gray, thick,dashed] (-0.75,-1.5) arc (-90:-270: 0.3 and 1.5 );
				\draw[black, thick,dashed] (-6,1.5) arc (90:-90: 0.3 and 1.5 );
				\draw[gray, thick,dashed] (-6,-1.5) arc (-90:-270: 0.3 and 1.5 );
				%
				% Blue shade for throat
				\filldraw [gray, thick,dashed, fill= blue!15, draw=none] (-6,0) circle [x radius=0.3, y radius=1.5];
				\filldraw[fill=blue, opacity = 0.3, draw = none] (-6,1.5) -- (-0.75,1.5) arc (90:-90: 0.3 and 1.5 ) -- (-6,-1.5) arc (-90:90: 0.3 and 1.5 ); 
				%
				%
				% Region delimiters...
				\draw [gray, thick,dashed] (1.5, -4) -- (1.5,0);
				%\draw [gray, thick,dotted] (1.5, 5) -- (1.5,0);
				\draw [gray, thick,dashed] (-0.75, -4) -- (-0.75,-1.5);
				\draw [gray, thick,dashed] (-6, -4) -- (-6,-1.5);
		        % ... and their coordinates
				\node[] at (-0.75, -4.5) {\Large $\rho=\rho_0(\lambda)$};
				\node[] at (-6, -4.5) {\Large $\rho=\rho_M$};
				\node[] at (1.5, -4.5) {\Large $\rho = 0$};
				%
				%
				% Asymptotics
				\draw[black, line width=1.5] (-6,-1.5) to[out=180, in=45] (-10, -4);
				\draw[black, line width=1.5] (-6,1.5) to[out=180, in=-45] (-10, 4);
				\draw [black, line width=1.5](-10,0) circle [x radius=0.4, y radius=4];
				%
				%
				% Names of the regions
				\node[align = center, centered] at (0.375, -3) {\Large Smooth\\\Large cap};
				\node[align = center, centered] at (-3.375, -3) {\Large AdS$_2\times S^3$ throat};
				\node[align = center, centered, fill=white] at (-8, -3) {\Large Asymptotic\\ \Large $\mathbb{R}^{1,3}\times S^1$ region };
				\node[align = center, centered] at (3.6, -3) {\Large Infinite\\ \Large AdS$_2 \times S^3$ \Large throat};
				%
				%
				% Centres
				\filldraw[red] (1,0.9) circle (2.5pt);
				\filldraw[red] (0.9,-1) circle (2.5pt);
				\filldraw[red] (0.6,0.1) circle (2.5pt);
				\draw[red, thick] (1,0.9) -- (0.9,-1);
				\draw[red, thick] (1,0.9) -- (0.6,0.1);
				\draw[red, thick] (0.9,-1) -- (0.6,0.1);
				%
				%
				% Arrow for the scaling limit
				\draw [ blue, line width=1.5, ->] (-0.75, 2.5) --(2,2.5);
				\draw [gray, thick,dashed] (-0.75,1.5) -- (-0.75,3.5);
				\node[align = center, centered] at (0.65, 3) {\Large \textcolor{blue}{$\lambda \rightarrow 0$}};
			\end{scope}

		\end{tikzpicture}
	\end{adjustbox}
	\caption{
		A schematic depiction of the bubbling geometries. The centres at coordinate $\lambda \, \Vec{d_j}$, depicted in red, lie in the cap region. In the scaling limit ($\lambda \rightarrow 0$), the geometry of the cap and of the asymptotic region remains fixed, while the throat region delimited by $\rho \in \[\rho_0(\lambda), \rho_M\]$ and shaded in blue, becomes longer and longer.
		}
\end{figure}
The throat region we are interested in is defined to be delimited by $\rho \in \[\rho_0(\lambda), \rho_M\]$. See Fig. \ref{fig:deep-throat_geometry}. 
The upper-bound, $\rho_M$, is chosen so that the top of our region of interest is inside the $\mathrm{AdS}_2$ throat: $\rho_M \ll Q_I$. The lower-bound, $\rho_0(\lambda)$, is chosen to be not too close from the centres, say $\rho_0(\lambda)=2 \lambda \max |\Vec{d_j}|$.
Then in the throat region, the metric is well-approximated by 
\begin{align} \label{eq:g_near-horizon}
    ds_{5}^2 = &- \frac{\rho^2}{Q_M^2} \(dt+\omega \)^2 
+  \frac{Q_M}{\rho^2} d\rho^2
+ Q_M d\Omega_3^2 \,,
\end{align}
where $Q_M\equiv\( Q_1Q_2Q_3 \)^{1/3}$ and
\begin{align}
 d\Omega_3^2   = \(d\psi + A\)^2 + d\theta^2 + \sin^2 \theta \, d\phi^2 \,, \qquad A\equiv\cos \theta d\phi \,.
\end{align}

In the basis $\cB\equiv \( dt+\omega, d\rho, d\psi + A, d\theta, \sin \theta \, d\phi \)$, the inverse metric is written in diagonal form:
\begin{align}
\label{eq:g_inverse}
g^{-1}=\mathrm{diag} \[ \frac{(Q_M)^2}{\rho^2}, \frac{\rho^2}{Q_M}, 
\frac{1}{Q_M}, \frac{1}{Q_M}, \frac{1}{Q_M} \] \,.
\end{align}

\textit{b. The derivatives in the throat region.}
Recall that
\begin{align}
\frac{\partial |\Vec{\rho}-\lambda \, \Vec{d_j}|}{\partial \lambda} = - \frac{(\Vec{\rho}-\lambda \, \Vec{d_j})\cdot \Vec{d_j}}{|\Vec{\rho}-\lambda \, \Vec{d_j}|} \,.
\end{align}
Then, given a harmonic function $H$ of the form (\ref{eq:harmonic_func_general}), the derivative of $H$ with respect to the scaling parameter $\lambda$ is given by
\begin{align}
\frac{\partial H}{\partial \lambda} = \sum_{j=1}^{n} h_j \frac{\( \Vec{\rho}-\lambda \, \Vec{d_j} \)\cdot \Vec{d_j}}{|\Vec{\rho}-\lambda \, \Vec{d_j}|^3} \,.
\end{align}
In the throat region, as $\rho \gg \lambda$, 
\begin{align}
\label{eq:dH/dlambda}
\frac{\partial H}{\partial \lambda} \sim \frac{\Vec{\rho}}{\rho^3} \cdot\(\sum_{j=1}^{n} h_j \Vec{d_j}\) = \cO\(1/\rho^2\)\,.
\end{align}
Thus, the derivative with respect to $\lambda$ of any harmonic function is an $\cO\(1/\rho^2\)$-function in the throat region.

Using (\ref{eq:dH/dlambda}) and that the harmonic functions are of order $\cO(1/\rho)$ in the throat region, we successively deduce, from formula (\ref{eq:Z_I_intermsof_harmonic}) and $Z_M\equiv\( Z_1Z_2Z_3 \)^{1/3}$, that
\begin{align}
\frac{\partial Z_I}{\partial \lambda}  = \cO\(1/\rho^2\)\,, \qquad
    \frac{\partial Z_M}{\partial \lambda}  = \cO\(1/\rho^2\)\,.
\end{align}
Written in the basis $\cB$, the derivative of the metric (\ref{eq:BubblingMetric}) with respect to $\lambda$ gives in the throat region
\begin{align} \label{eq:dg/dlambda}
\frac{\partial g}{\partial \lambda} = \mathrm{diag} \[ 
\cO(\rho), \cO(1/\rho^3), \cO(1/\rho), \cO(1/\rho),\cO(1/\rho)  \] \,.
\end{align}
%\YL{Transverse traceless}

\textit{c. The distance to the scaling limit.}
Combining (\ref{eq:g_inverse}) and (\ref{eq:dg/dlambda}) gives that in the basis $\cB$,
\begin{align}
\frac{\partial g}{\partial \lambda} g^{-1} = \mathrm{diag} \[ \cO(1/\rho), \cO(1/\rho), \cO(1/\rho), \cO(1/\rho), \cO(1/\rho) \] \,,
\end{align}
so that
\begin{align} \label{eq:trace_1/rho^2}
    \tr\[ \( \frac{\partial g}{\partial \lambda} g^{-1} \)^2 \] = \cO(1/\rho^2) \,.
\end{align}

Therefore, the metric component on moduli space behaves like
\begin{align}
\int_{\cM} \sqrt{g} \operatorname{tr}\left[\left(g^{-1} \frac{\partial g}{\partial \tau}\right)^{2}\right]
\propto \int_{\rho_0(\lambda)}^{\rho_M} d\rho \,\frac{1}{\rho^2}
\propto \frac{1}{\lambda} \,.
\end{align}
In the scaling limit, the metric component on moduli space along $\lambda$ blows up like $1/\lambda$, but this gives a DeWitt distance (\ref{eq:DeWitt_distance_without_volume}) (between a point $p(\lambda_0)$ in bulk moduli space and a point $p(\lambda)$ closer to the scaling limit)
\begin{align} \label{eq:distance_scaling_generalized}
    \Delta_\mathrm{DeWitt\,1}(\lambda_0,\lambda) \propto c\int_{\lambda}^{\lambda_0} d\lambda' \, \frac{1}{\sqrt{\lambda'}} \propto \sqrt{\lambda_0}-\sqrt{\lambda} \,,
\end{align}
which is \textit{finite}. \emph{This dependence in the scaling parameter, $\lambda$, matches exactly that of the phase-space distance (\ref{eq:Delta_phase_scaling})!}
\newline

Note that if we wanted $\Delta_{\mathrm{DeWitt\,1}}$ to match the infinite distance of $\Delta_{\textrm{exponential}}\sim |L_{\mathrm{throat},i}-L_{\mathrm{throat},f}|$, we would have needed
\begin{align} \label{eq:DeWitt_false}
\Delta_{\mathrm{DeWitt\,1},\, \mathrm{false}} \propto -\ln\(\frac{\lambda_i}{\lambda_f}\) \,,
\end{align}
that is to say
\begin{align}
    \tr\[ \( \frac{\partial g}{\partial \lambda} g^{-1} \)^2 \] \propto \frac{1}{\rho^3} \,.
\end{align}
This last behaviour is too singular and impossible given (\ref{eq:trace_1/rho^2}), whose most singular power in $\rho$ near $\rho=0$ is \textit{at most} $1/\rho^2$.

\textit{d. The distance with the volume factor.}
%
%From (\ref{eq:g_near-horizon}) that $\sqrt{g}$ is constant with respect to $\rho$ in the throat region.
%So the volume of the throat region delimited by $\rho \in \[\rho_0(\lambda), \rho_M\]$ is proportional to $\rho_M-\rho_0(\lambda) \approx \rho_M$, which stays constant in the scaling limit.
The volume of the asymptotic $\mathbb{R}^{1,3}\times S^1$ region is infinite, so in the chosen coordinates, the DeWitt distance with the volume factor (\ref{eq:generalized_distance}) between any two bubbling geometries is 0.

In fact, the distance (\ref{eq:generalized_distance}) is not invariant under diffeomorphisms of the spacetime metric, and, as explained in \cite{Bonnefoy:2019nzv}, one should choose a frame which satisfies the condition of a vanishing Lie derivative. For metrics parameterized by a single dimensionfull scale, the distance (\ref{eq:generalized_distance}) computed in such a frame gives a logarithmic divergence \cite{Bonnefoy:2019nzv}. Thus, for a bubbling geometry approaching the scaling limit and only parameterized by the scaling parameter $\lambda$, the distance to the scaling limit diverges logarithmically:
\begin{align} \label{eq:DeWitt_2_scaling_limit}
\Delta_\mathrm{DeWitt\,2} \propto -\ln\(\frac{\lambda_i}{\lambda_f}\) \,.
\end{align}
\emph{This behaviour in $\lambda$ matches that of the exponential distance (\ref{eq:exponential_distance}).}

%%%%%%%%%%%%%%%%%%%%%%%%%%%%%%%%%%%%%%%%%%%%%%%%%%%%%%%%%%%%%%%%%%%%%%%%%%%%%%%%%%%%%%%%%%%%%%%%%%
%%%%%%%%%%%%%%%%%%%%%%%%%%%%%%%%%%%%%%%%%%%%%%%%%%%%%%%%%%%%%%%%%%%%%%%%%%%%%%%%%%%%%%%%%%%%%%%%%%
%%%% 	Discussion and conclusion
%%%%%%%%%%%%%%%%%%%%%%%%%%%%%%%%%%%%%%%%%%%%%%%%%%%%%%%%%%%%%%%%%%%%%%%%%%%%%%%%%%%%%%%%%%%%%%%%%%
%%%%%%%%%%%%%%%%%%%%%%%%%%%%%%%%%%%%%%%%%%%%%%%%%%%%%%%%%%%%%%%%%%%%%%%%%%%%%%%%%%%%%%%%%%%%%%%%%%
\section{The conflict over moduli space}
\label{sec:discussion}

In this letter, we have compared the distance in moduli space between two bubbling geometries of different throat lengths, using the two DeWitt distances. 
Our computation in Section \ref{sec:computation} shows an agreement between the distance to the scaling limit according to the phase-space distance (\ref{eq:phase_distance}) computed in \cite{Li:2021gbg} and to the DeWitt distance \textit{without the volume factor} (\ref{eq:DeWitt_distance_without_volume}). Both formulas, (\ref{eq:Delta_phase_scaling}) and (\ref{eq:DeWitt_distance_without_volume}), give a finite result for this distance.
On the other hand, the distance to the scaling limit according to the DeWitt distance \textit{with the volume factor} (\ref{eq:generalized_distance}) is infinite and matches with that of the exponential distance (\ref{eq:exponential_distance}). %There, a tower of massless states emerges at infinite distance in the moduli space.

One of the motivations to take (\ref{eq:generalized_distance}) to be the generalised Swampland distance is that it gives the exponential mass decrease (\ref{eq:Swampland_distance_conjecture}) with respect to the moduli space distance \cite{Bonnefoy:2019nzv}. But the distance (\ref{eq:DeWitt_distance_without_volume}) is also a well-defined distance that can be used in the context of the Swampland.
The two DeWitt distances (\ref{eq:DeWitt_distance_without_volume}) and (\ref{eq:generalized_distance}) do not agree, and measure different physical notions -- if they correspond to any physical notions at all. Could the lew-velocity distance match one of the two DeWitt distances?
%\newline
\vspace{1em}

At first sight, both the results (\ref{eq:distance_scaling_generalized}) and (\ref{eq:DeWitt_2_scaling_limit}) appear to be in tension with an earlier works in the literature \cite{Maldacena:1998uz,Denef:2002ru}, which use the low-velocity distance (\ref{eq:low-velocity_distance}) of \cite{Ferrell:1987gf}.

Our computation in Section \ref{sec:computation} leading to (\ref{eq:distance_scaling_generalized}) did not involve the use of the smoothness conditions (\ref{eq:smoothness_cond_GH}) nor the bubble equations (\ref{eq:BubbleEquations}). Indeed, we applied the DeWitt distance to $\mathrm{AdS}_2$ throat regions of different lengths. As the throat is far from the centres, it is insensitive to the physics at the bottom of the throat. For instance, the throat is insensitive to the details of the charges $\mathbf{h}_j$ at the centres, as long as they give the right black-hole asymptotic charges (\ref{eq:Asymptotic_charges_5d}). 

As such, one can take $k_j^I=0$ for all $j$ and $I$, so that the black-hole charge $Q_I$ is only given by the $l_j^I$'s: 
\begin{align}
Q_I = \sum_{j=1}^n l_j^I \,, \qquad Z_I=L_I \,,
\end{align}
and one would still find the same finite result for the distance to the scaling limit (\ref{eq:distance_scaling_generalized}).
In addition to the condition $K^I=0$, let us take the simple example of a two-center solution and further impose
\begin{align} \label{eq:harmonic_func_to_BH}
V=L_1=L_2=L_3  \,.
\end{align}

From the four-dimensional point of view, equation (\ref{eq:harmonic_func_to_BH}) means that the centres of the multi-centered solutions become extremal Reissner-Nordstr\"om black holes at the same coordinate location.
However, this is exactly the system considered in \cite{Maldacena:1998uz} and previously in \cite{Ferrell:1987gf} (see also \cite{Michelson:1997vq}). In \cite{Maldacena:1998uz}, it was computed that the low-velocity metric in moduli space is
\begin{align}
ds^2 = \frac{1}{2}\frac{(l_1)^3l_2+l_1(l_2)^3}{(\rho_{12})^3}\, (\mathrm{d}\rho_{12})^2 \,,
\end{align}
where $\rho_{12}$ denotes the distance between centre 1 and 2, and $l_1$, $l_2$ are the respective electric charges at centre 1 and 2. (The results match the same $1/\rho^3$ dependence found in \cite{Denef:2002ru}.)
Therefore, the distance to the scaling limit is given by
\begin{align} \label{eq:low-v_to_scaling}
\Delta_{\mathrm{low-v.}}(\lambda_0,\lambda) \propto \int_{\lambda}^{\lambda_0} d\lambda \frac{1}{\lambda^{3/2}}
\propto \frac{1}{\lambda^{1/2}} - \frac{1}{{\lambda_0}^{1/2}} \,.
\end{align}
The distance to the scaling limit is infinite, and blows up like $\frac{1}{\lambda^{1/2}}$. This is, at first sight, in contradiction not only with the distance to the scaling limit according to the phase-space distance (\ref{eq:Delta_phase_scaling}) and to the first DeWitt distance (\ref{eq:distance_scaling_generalized}), but also with the second DeWitt distance (\ref{eq:DeWitt_2_scaling_limit}).

However, there is a subtlety here:%
\footnote{I would like to thank Micha Berkooz for pointing out this subtlety.}
the low-velocity distance measures the moduli-space distance on a higher dimensional moduli space. For instance, the dimension of three-centered solutions before applying the bubble equations (\ref{eq:BubbleEquations}) is 6, while after imposing them the dimension becomes 4. The low-velocity distance measures distances on the 6-dimensional moduli space, while the phase-space distance and the two DeWitt distances measure the distance on the 4-dimensional moduli space. This is why the low-velocity distance does not necessarily contradict both the DeWitt distances.
%\footnote{Besides, note that this infinite distance does not match with the exponential distance (\ref{eq:exponential_distance}) -- which has a logarithmic divergence in $\lambda$ (\ref{eq:DeWitt_false}) -- either.}

%The low-velocity distance to the scaling limit (\ref{eq:low-v_to_scaling}), although infinite, does not match with the exponential distance (\ref{eq:exponential_distance}) -- which has a logarithmic divergence in $\lambda$ (\ref{eq:DeWitt_false}) -- either. Indeed, its $\frac{1}{\lambda^{1/2}}$-divergence would lead to a tower of KK states whose mass is decreasing like
%\begin{align}
%    M(\lambda) \sim \frac{1}{(\Delta_{\mathrm{low-v.}})^2} \,,
%\end{align}
%instead of the exponential decrease (\ref{eq:Swampland_distance_conjecture}) in the usual Swampland picture, if the converse of the Generalized Distance Conjecture applies in this limit.

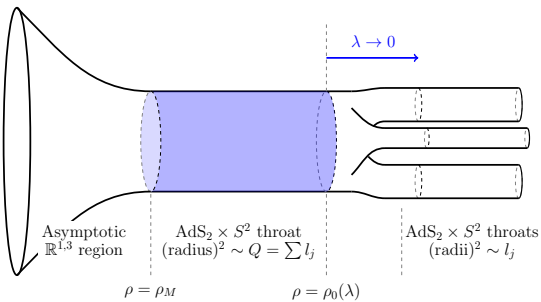
\begin{figure}[h]
	\label{fig:multi-centered_black_holes}
	\begin{adjustbox}{max totalsize={0.4\textwidth}{\textheight},center}
		\begin{tikzpicture}
			\begin{scope}[shift={(0,0)}]
			    % Cap
				%\draw[black,line width=1.5] (0, 1.5) arc (90:-90:1.5);
				%\draw [gray, thick,dashed](0.0,0) circle [x radius=0.3, y radius=1.5];
				% Throat
				\draw [ black, line width=1.5] (0, 1.5) --(-6,1.5);
				\draw [ black, line width=1.5] (0, -1.5) --(-6,-1.5);
				%\draw [gray, thick,dashed, ] (-0.75,0) circle [x radius=0.3, y radius=1.5];
				\draw[black, thick,dashed] (-0.75,1.5) arc (90:-90: 0.3 and 1.5 );
				\draw[gray, thick,dashed] (-0.75,-1.5) arc (-90:-270: 0.3 and 1.5 );
				% Infinite throat
				%\draw [gray, thick, dashed] (0,1.5) -- (5, 1.5);
				%\draw [gray, thick, dashed] (0,-1.5) -- (5, -1.5);
				%\draw [gray, thick,dashed, ] (3,0) circle [x radius=0.3, y radius=1.5];
				% Blue shade for throat
				%\filldraw [gray, thick,dashed, fill= blue!15] (-0.75,0) circle [x radius=0.3, y radius=1.5];
				%\filldraw[fill=blue, opacity = 0.3, draw = none] (-6,1.5) -- (-0.75,1.5) arc (90:-90: 0.3 and 1.5 ) -- (-6,-1.5) arc (-90:90: 0.3 and 1.5 ); 
				\filldraw [gray, thick,dashed, fill= blue!15, draw=none] (-6,0) circle [x radius=0.3, y radius=1.5];
				\filldraw[fill=blue, opacity = 0.3, draw = none] (-6,1.5) -- (-0.75,1.5) arc (90:-90: 0.3 and 1.5 ) -- (-6,-1.5) arc (-90:90: 0.3 and 1.5 ); 
				\draw[black, thick,dashed] (-6,1.5) arc (90:-90: 0.3 and 1.5 );
				\draw[gray, thick,dashed] (-6,-1.5) arc (-90:-270: 0.3 and 1.5 );
				%
				%
				% Region delimiters...
				\draw [gray, thick,dashed] (1.5, -4) -- (1.5,-1.9);
				\draw [gray, thick,dashed] (-0.75, -4) -- (-0.75,-1.5);
				\draw [gray, thick,dashed] (-6, -4) -- (-6,-1.5);
		        % ... and their coordinates
				\node[] at (-0.75, -4.5) {\Large $\rho=\rho_0(\lambda)$};
				\node[] at (-6, -4.5) {\Large $\rho=\rho_M$};
				%\node[] at (1.5, -4.5) {\Large $\rho = 0$};
				%
				%
				% Asymptotics
				\draw[black, line width=1.5] (-6,-1.5) to[out=180, in=45] (-10, -4);
				\draw[black, line width=1.5] (-6,1.5) to[out=180, in=-45] (-10, 4);
				%\draw [gray, thick,dashed](-10,0) circle [x radius=0.4, y radius=4];
				\draw [black, line width=1.5](-10,0) circle [x radius=0.4, y radius=4];
				%
				%
				% Individual black holes
				\draw [gray, thick,dashed, ] (5,1.1) circle [x radius=0.1, y radius=0.5];
				\draw [gray, thick,dashed, ] (5,-1.2) circle [x radius=0.1, y radius=0.5];
				\draw [gray, thick,dashed, ] (5.25,0.1) circle [x radius=0.07, y radius=0.3];
				%
				%\draw [gray, thick,dashed, ] (2,1.1) circle [x radius=0.1, y radius=0.5];
				\draw[black, thick,dashed] (2,1.6) arc (90:-90: 0.1 and 0.5 );
				\draw[gray, thick,dashed] (2,0.6) arc (-90:-270: 0.1 and 0.5 );
				%\draw [gray, thick,dashed, ] (2,-1.2) circle [x radius=0.1, y radius=0.5];
				\draw[black, thick,dashed] (2,-0.7) arc (90:-90: 0.1 and 0.5 );
				\draw[gray, thick,dashed] (2,-1.7) arc (-90:-270: 0.1 and 0.5 );
				%\draw [gray, thick,dashed, ] (2.25,0.1) circle [x radius=0.07, y radius=0.3];
				\draw[black, thick,dashed] (2.25,0.4) arc (90:-90: 0.07 and 0.3 );
				\draw[gray, thick,dashed] (2.25,-0.2) arc (-90:-270: 0.07 and 0.3 );
				% three plain arcs at the end
				\draw[black, line width=1.] (5,1.6) arc (90:-90: 0.1 and 0.5 );
				\draw[black, line width=1.] (5,-0.7) arc (90:-90: 0.1 and 0.5 );
				\draw[black, line width=1.] (5.25,0.4) arc (90:-90: 0.07 and 0.3 );
				\draw [ black, line width=1.5] (1, 1.6) --(5,1.6);
				\draw [ black, line width=1.5] (1, 0.6) --(5,0.6);
				\draw [ black, line width=1.5] (1, -1.7) --(5,-1.7);
				\draw [ black, line width=1.5] (1, -0.7) --(5,-0.7);
				\draw [ black, line width=1.5] (1, 0.4) --(5.25,0.4);
				\draw [ black, line width=1.5] (1, -0.2) --(5.25,-0.2);
				%
				%\draw[black, line width=1.5] (1, 1.6) to[out=180, in=0] (0, 1.5);
				%
				%
				% Merging region
				%\draw[black,line width=1.5] (1, -0.7) arc (270:90:0.65);
				\draw[black, line width=1.5] (1, 1.6) to[out=180, in=0] (0, 1.5);
				\draw[black, line width=1.5] (1, -1.7) to[out=180, in=0] (0, -1.5);
				\draw[black, line width=1.5] (1, 0.4) to[out=180, in=-45] (0, 1);
				\draw[black, line width=1.5] (1, -0.2) to[out=180, in=45] (0, -1);
				\draw[black, line width=1.5] (1, -0.7) to[out=180, in=-45] (0.5, -0.45);
				\draw[black, line width=1.5] (0.65, 0.45) to[out=45, in=180] (1, 0.6);
				%
				%
				% Names of the regions
				%\node[align = center, centered] at (0.375, -3) {\Large Smooth\\\Large cap};
				%
				\node[align = center, centered] at (-3.375, -3) {\Large AdS$_2\times S^2$ throat\\ \Large (radius)$^2$ $\sim Q= \sum l_j$};
				\node[align = center, centered, fill=white] at (-8, -3) {\Large Asymptotic\\ \Large $\mathbb{R}^{1,3}$ region };
				\node[align = center, centered] at (3.6, -3) {\Large AdS$_2\times S^2$ throats\\ \Large (radii)$^2$ $\sim l_j$};
				%
				%
				% Centres
				%\filldraw[red] (4,1.1) circle (2.5pt);
				%\filldraw[red] (4,-1.2) circle (2.5pt);
				%\filldraw[red] (4.25,0.1) circle (2.5pt);
				%\draw[red, thick] (1,0.9) -- (0.9,-1);
				%\draw[red, thick] (1,0.9) -- (0.6,0.1);
				%\draw[red, thick] (0.9,-1) -- (0.6,0.1);
				%
				%
				% Arrow for the scaling limit
				\draw [ blue, line width=1.5, ->] (-0.75, 2.5) --(2,2.5);
				\draw [gray, thick,dashed] (-0.75,1.5) -- (-0.75,3.5);
				\node[align = center, centered] at (0.65, 3) {\Large \textcolor{blue}{$\lambda \rightarrow 0$}};
			\end{scope}
		\end{tikzpicture}
	\end{adjustbox}
	\caption{
		Geometry of multiple near-coincident extremal black holes. The main AdS$_2$ throat, shaded in blue, divides itself into multiple infinite AdS$_2$ throats corresponding to the individual black holes. In the scaling limit ($\lambda \rightarrow 0$), only the length of the main throat increases.
		}
\end{figure}
But regardless of the above subtlety, the physical picture of the scaling limit of multiple extremal black holes is quite similar to that of the smooth cap being pushed deeper by the $\mathrm{AdS}_2$ throat. For near-coincident extremal black holes, after the asymptotically flat region, there is a main $\mathrm{AdS}_2$ throat region corresponding to the sum of all the black hole charges, $Q = \sum_{j=1}^n l_j$. See Fig. \ref{fig:multi-centered_black_holes}. This main throat region eventually divides itself into multiple $\mathrm{AdS}_2$ throat regions, corresponding to the different local charges, $l_j$, of the individual extremal black holes \cite{Maldacena:1998uz}. In the scaling limit, the geometry of the bottom of the throat remains fixed, and only the main $\mathrm{AdS}_2$ throat of the bigger black hole becomes longer and longer \cite{Michelson:1999dx,Maldacena:1998uz}. 
Thus, the main $\mathrm{AdS}_2$ throat is insensitive to whether lies at its bottom a smooth cap or a set of black holes: this is why the computations in Section \ref{sec:computation} leading to (\ref{eq:distance_scaling_generalized}) will give the same result for a deepening $\mathrm{AdS}_2$ throat with black holes at its bottom.
%
%\footnote{A possible resolution of this tension could be that while the coordinates we chose are fine to compute the DeWitt distance for smooth geometries, one cannot use them to compute the DeWitt distance if horizons appear at the bottom of the throat \cite{Bonnefoy:2019nzv}. One might need to find some equivalent of Eddington-Finkelstein coordinates adapted to multiple black holes.}
\newline

In this letter, we compared the phase-space distance and the two DeWitt distances, on a moduli space of geometries developing longer and longer $\mathrm{AdS}_2$ throats. 
In the specific example of path on moduli space we took, the phase-space distance and the DeWitt distance \textit{without the volume factor} agree, while the DeWitt distance \textit{with the volume factor} disagrees with them:
%\vspace{-\topsep}
\begin{itemize}
\setlength{\itemsep}{1pt plus 1pt}
\item According to the phase-space distance and the DeWitt distance \textit{without the volume factor}, the locus in moduli space where the lengths of these throats become infinite -- \textit{i.e.} the \textit{scaling limit} locus -- lies at \textit{finite distance} in moduli space. Indeed, for both of these distances, the metric component on moduli space along the scaling parameter, $\lambda$, blows up as $1/\lambda$ in the vicinity of the scaling limit ($\lambda \rightarrow 0$), so the distance to the scaling limit is finite.\\
In the scaling limit, an infinite tower of KK modes becomes massless due to the increasing redshift from the cap to the asymptotics. According to the phase-space distance and the DeWitt distance \textit{without the volume factor}, this infinite tower of massless states emerges at finite distance in moduli space. If the DeWitt distance \textit{without the volume factor} is the relevant distance to use in the context of the Swampland, these results constitute a counter-example to the converse of the Generalized Distance Conjecture. 
\item  However, according to the DeWitt distance \textit{with the volume factor}, the distance in moduli space to the scaling limit locus is \textit{infinite}, and scales like $\log \lambda$. This $\log \lambda$ behaviour would lead to a tower of KK states whose mass decrease matches the exponential decrease of the usual Swampland picture, should the converse of the Generalized Distance Conjecture apply in the scaling limit of deep-throat geometries.
%the logarithmic growth in $1/\lambda$ one would expect if the converse of the Generalized Distance Conjecture applies in this limit.
%There, the metric component along the scaling parameter blows up like
%
%\item According to the low-velocity distance, the distance in moduli space to the scaling limit locus is \textit{infinite}, and scales like ${1}/{\lambda^{1/2}}$. However, the low-velocity distance computed here concerns a \textit{higher-dimensional moduli space of solutions} on which one has not yet imposed the bubbles equations. Finding out which are the fields that need to be integrated out in order to get the lower-dimensional moduli space will indicate the behaviour of the low-velocity distance to the scaling limit for bubbling geometries. This last result should match one of the two DeWitt distances, and thus shedding light on the physical interpretation of the matched one. We hope to report on this in the future.
\end{itemize}

According to the low-velocity distance, the distance in moduli space to the scaling limit locus is \textit{infinite}, and scales like ${1}/{\lambda^{1/2}}$. However, the low-velocity distance computed here concerns a \textit{higher-dimensional moduli space of solutions} on which one has not yet imposed the bubbles equations. Finding out which are the fields that need to be integrated out in order to get the lower-dimensional moduli space will indicate the behaviour of the low-velocity distance to the scaling limit for bubbling geometries. This last result should match one of the two DeWitt distances, and thus shedding light on the physical interpretation of the matched distance. We hope to report on this in the future.
%\vspace{\baselineskip}%\linebreak
%\vspace{1cm}
\vspace*{\baselineskip}

\textbf{Acknowledgements.} I would like to thank Iosif Bena, Micha Berkooz, Jan de Boer, Anthony Houppe and Dieter L\"ust for useful discussions.
This work was partially supported
by the ERC Grants ``772408 - Stringlandscape'' and ``787320 - QBH Structure''. %, the ANR grant Black-dS-String ANR-16-CE31-0004-01 and the John Templeton Foundation grant 61169.

%%%%%%%%%%%%%%%%%%%%%%%%%%%%%%%%%%%%%%%%%%%%%%%%%%
%\clearpage

%\bibliographystyle{JHEP}

\bibliography{TowersFiniteDistance}

\begin{thebibliography}{36}%
\makeatletter
\providecommand \@ifxundefined [1]{%
 \@ifx{#1\undefined}
}%
\providecommand \@ifnum [1]{%
 \ifnum #1\expandafter \@firstoftwo
 \else \expandafter \@secondoftwo
 \fi
}%
\providecommand \@ifx [1]{%
 \ifx #1\expandafter \@firstoftwo
 \else \expandafter \@secondoftwo
 \fi
}%
\providecommand \natexlab [1]{#1}%
\providecommand \enquote  [1]{``#1''}%
\providecommand \bibnamefont  [1]{#1}%
\providecommand \bibfnamefont [1]{#1}%
\providecommand \citenamefont [1]{#1}%
\providecommand \href@noop [0]{\@secondoftwo}%
\providecommand \href [0]{\begingroup \@sanitize@url \@href}%
\providecommand \@href[1]{\@@startlink{#1}\@@href}%
\providecommand \@@href[1]{\endgroup#1\@@endlink}%
\providecommand \@sanitize@url [0]{\catcode `\\12\catcode `\$12\catcode
  `\&12\catcode `\#12\catcode `\^12\catcode `\_12\catcode `\%12\relax}%
\providecommand \@@startlink[1]{}%
\providecommand \@@endlink[0]{}%
\providecommand \url  [0]{\begingroup\@sanitize@url \@url }%
\providecommand \@url [1]{\endgroup\@href {#1}{\urlprefix }}%
\providecommand \urlprefix  [0]{URL }%
\providecommand \Eprint [0]{\href }%
\providecommand \doibase [0]{http://dx.doi.org/}%
\providecommand \selectlanguage [0]{\@gobble}%
\providecommand \bibinfo  [0]{\@secondoftwo}%
\providecommand \bibfield  [0]{\@secondoftwo}%
\providecommand \translation [1]{[#1]}%
\providecommand \BibitemOpen [0]{}%
\providecommand \bibitemStop [0]{}%
\providecommand \bibitemNoStop [0]{.\EOS\space}%
\providecommand \EOS [0]{\spacefactor3000\relax}%
\providecommand \BibitemShut  [1]{\csname bibitem#1\endcsname}%
\let\auto@bib@innerbib\@empty
%</preamble>
\bibitem [{\citenamefont {Ritter}(2002)}]{Ritter:2002zg}%
  \BibitemOpen
  \bibfield  {author} {\bibinfo {author} {\bibfnamefont {W.~G.}\ \bibnamefont
  {Ritter}},\ }\href@noop {} {\  (\bibinfo {year} {2002})},\ \Eprint
  {http://arxiv.org/abs/math-ph/0208008} {arXiv:math-ph/0208008} \BibitemShut
  {NoStop}%
\bibitem [{\citenamefont {Echeverria-Enriquez}\ \emph
  {et~al.}(1998)\citenamefont {Echeverria-Enriquez}, \citenamefont
  {Munoz-Lecanda}, \citenamefont {Roman-Roy},\ and\ \citenamefont
  {Victoria-Monge}}]{Echeverria-Enriquez:1998umj}%
  \BibitemOpen
  \bibfield  {author} {\bibinfo {author} {\bibfnamefont {A.}~\bibnamefont
  {Echeverria-Enriquez}}, \bibinfo {author} {\bibfnamefont {M.~C.}\
  \bibnamefont {Munoz-Lecanda}}, \bibinfo {author} {\bibfnamefont
  {N.}~\bibnamefont {Roman-Roy}}, \ and\ \bibinfo {author} {\bibfnamefont
  {C.}~\bibnamefont {Victoria-Monge}},\ }\href@noop {} {\bibfield  {journal}
  {\bibinfo  {journal} {Extracta Math.}\ }\textbf {\bibinfo {volume} {13}},\
  \bibinfo {pages} {135} (\bibinfo {year} {1998})},\ \Eprint
  {http://arxiv.org/abs/math-ph/9904008} {arXiv:math-ph/9904008} \BibitemShut
  {NoStop}%
\bibitem [{\citenamefont {Berman}\ and\ \citenamefont
  {Cardoso}(2022)}]{Berman:2022acl}%
  \BibitemOpen
  \bibfield  {author} {\bibinfo {author} {\bibfnamefont {D.~S.}\ \bibnamefont
  {Berman}}\ and\ \bibinfo {author} {\bibfnamefont {G.}~\bibnamefont
  {Cardoso}},\ }\href@noop {} {\  (\bibinfo {year} {2022})},\ \Eprint
  {http://arxiv.org/abs/2201.00349} {arXiv:2201.00349 [hep-th]} \BibitemShut
  {NoStop}%
\bibitem [{\citenamefont {Maoz}\ and\ \citenamefont
  {Rychkov}(2005)}]{Maoz:2005nk}%
  \BibitemOpen
  \bibfield  {author} {\bibinfo {author} {\bibfnamefont {L.}~\bibnamefont
  {Maoz}}\ and\ \bibinfo {author} {\bibfnamefont {V.~S.}\ \bibnamefont
  {Rychkov}},\ }\href {\doibase 10.1088/1126-6708/2005/08/096} {\bibfield
  {journal} {\bibinfo  {journal} {JHEP}\ }\textbf {\bibinfo {volume} {08}},\
  \bibinfo {pages} {096} (\bibinfo {year} {2005})},\ \Eprint
  {http://arxiv.org/abs/hep-th/0508059} {arXiv:hep-th/0508059} \BibitemShut
  {NoStop}%
\bibitem [{\citenamefont {Rychkov}(2006)}]{Rychkov:2005ji}%
  \BibitemOpen
  \bibfield  {author} {\bibinfo {author} {\bibfnamefont {V.~S.}\ \bibnamefont
  {Rychkov}},\ }\href {\doibase 10.1088/1126-6708/2006/01/063} {\bibfield
  {journal} {\bibinfo  {journal} {JHEP}\ }\textbf {\bibinfo {volume} {01}},\
  \bibinfo {pages} {063} (\bibinfo {year} {2006})},\ \Eprint
  {http://arxiv.org/abs/hep-th/0512053} {arXiv:hep-th/0512053} \BibitemShut
  {NoStop}%
\bibitem [{\citenamefont {de~Boer}\ \emph {et~al.}(2009)\citenamefont
  {de~Boer}, \citenamefont {El-Showk}, \citenamefont {Messamah},\ and\
  \citenamefont {Van~den Bleeken}}]{deBoer:2008zn}%
  \BibitemOpen
  \bibfield  {author} {\bibinfo {author} {\bibfnamefont {J.}~\bibnamefont
  {de~Boer}}, \bibinfo {author} {\bibfnamefont {S.}~\bibnamefont {El-Showk}},
  \bibinfo {author} {\bibfnamefont {I.}~\bibnamefont {Messamah}}, \ and\
  \bibinfo {author} {\bibfnamefont {D.}~\bibnamefont {Van~den Bleeken}},\
  }\href {\doibase 10.1088/1126-6708/2009/05/002} {\bibfield  {journal}
  {\bibinfo  {journal} {JHEP}\ }\textbf {\bibinfo {volume} {05}},\ \bibinfo
  {pages} {002} (\bibinfo {year} {2009})},\ \Eprint
  {http://arxiv.org/abs/0807.4556} {arXiv:0807.4556 [hep-th]} \BibitemShut
  {NoStop}%
\bibitem [{\citenamefont {Mayerson}\ and\ \citenamefont
  {Shigemori}(2021)}]{Mayerson:2020acj}%
  \BibitemOpen
  \bibfield  {author} {\bibinfo {author} {\bibfnamefont {D.~R.}\ \bibnamefont
  {Mayerson}}\ and\ \bibinfo {author} {\bibfnamefont {M.}~\bibnamefont
  {Shigemori}},\ }\href {\doibase 10.21468/SciPostPhys.10.1.018} {\bibfield
  {journal} {\bibinfo  {journal} {SciPost Phys.}\ }\textbf {\bibinfo {volume}
  {10}},\ \bibinfo {pages} {018} (\bibinfo {year} {2021})},\ \Eprint
  {http://arxiv.org/abs/2010.04172} {arXiv:2010.04172 [hep-th]} \BibitemShut
  {NoStop}%
\bibitem [{\citenamefont {Witten}(1986)}]{Witten:1986qs}%
  \BibitemOpen
  \bibfield  {author} {\bibinfo {author} {\bibfnamefont {E.}~\bibnamefont
  {Witten}},\ }\href {\doibase 10.1016/0550-3213(86)90298-1} {\bibfield
  {journal} {\bibinfo  {journal} {Nucl. Phys. B}\ }\textbf {\bibinfo {volume}
  {276}},\ \bibinfo {pages} {291} (\bibinfo {year} {1986})}\BibitemShut
  {NoStop}%
\bibitem [{\citenamefont {Crnkovic}(1988)}]{Crnkovic:1987tz}%
  \BibitemOpen
  \bibfield  {author} {\bibinfo {author} {\bibfnamefont {C.}~\bibnamefont
  {Crnkovic}},\ }\href {\doibase 10.1088/0264-9381/5/12/008} {\bibfield
  {journal} {\bibinfo  {journal} {Class. Quant. Grav.}\ }\textbf {\bibinfo
  {volume} {5}},\ \bibinfo {pages} {1557} (\bibinfo {year} {1988})}\BibitemShut
  {NoStop}%
\bibitem [{\citenamefont {Ferrell}\ and\ \citenamefont
  {Eardley}(1987)}]{Ferrell:1987gf}%
  \BibitemOpen
  \bibfield  {author} {\bibinfo {author} {\bibfnamefont {R.~C.}\ \bibnamefont
  {Ferrell}}\ and\ \bibinfo {author} {\bibfnamefont {D.~M.}\ \bibnamefont
  {Eardley}},\ }\href {\doibase 10.1103/PhysRevLett.59.1617} {\bibfield
  {journal} {\bibinfo  {journal} {Phys. Rev. Lett.}\ }\textbf {\bibinfo
  {volume} {59}},\ \bibinfo {pages} {1617} (\bibinfo {year}
  {1987})}\BibitemShut {NoStop}%
\bibitem [{\citenamefont {Manton}(1982)}]{MANTON198254}%
  \BibitemOpen
  \bibfield  {author} {\bibinfo {author} {\bibfnamefont {N.}~\bibnamefont
  {Manton}},\ }\href {\doibase https://doi.org/10.1016/0370-2693(82)90950-9}
  {\bibfield  {journal} {\bibinfo  {journal} {Physics Letters B}\ }\textbf
  {\bibinfo {volume} {110}},\ \bibinfo {pages} {54} (\bibinfo {year}
  {1982})}\BibitemShut {NoStop}%
\bibitem [{\citenamefont {Michelson}\ and\ \citenamefont
  {Strominger}(1999)}]{Michelson:1999dx}%
  \BibitemOpen
  \bibfield  {author} {\bibinfo {author} {\bibfnamefont {J.}~\bibnamefont
  {Michelson}}\ and\ \bibinfo {author} {\bibfnamefont {A.}~\bibnamefont
  {Strominger}},\ }\href {\doibase 10.1088/1126-6708/1999/09/005} {\bibfield
  {journal} {\bibinfo  {journal} {JHEP}\ }\textbf {\bibinfo {volume} {09}},\
  \bibinfo {pages} {005} (\bibinfo {year} {1999})},\ \Eprint
  {http://arxiv.org/abs/hep-th/9908044} {arXiv:hep-th/9908044 [hep-th]}
  \BibitemShut {NoStop}%
%%CITATION = HEP-TH/9908044;%%
\bibitem [{\citenamefont {Douglas}\ \emph {et~al.}(1997)\citenamefont
  {Douglas}, \citenamefont {Polchinski},\ and\ \citenamefont
  {Strominger}}]{Douglas:1997vu}%
  \BibitemOpen
  \bibfield  {author} {\bibinfo {author} {\bibfnamefont {M.~R.}\ \bibnamefont
  {Douglas}}, \bibinfo {author} {\bibfnamefont {J.}~\bibnamefont {Polchinski}},
  \ and\ \bibinfo {author} {\bibfnamefont {A.}~\bibnamefont {Strominger}},\
  }\href {\doibase 10.1088/1126-6708/1997/12/003} {\bibfield  {journal}
  {\bibinfo  {journal} {JHEP}\ }\textbf {\bibinfo {volume} {12}},\ \bibinfo
  {pages} {003} (\bibinfo {year} {1997})},\ \Eprint
  {http://arxiv.org/abs/hep-th/9703031} {arXiv:hep-th/9703031} \BibitemShut
  {NoStop}%
\bibitem [{\citenamefont {Kaplan}\ and\ \citenamefont
  {Michelson}(1997)}]{Kaplan:1997gk}%
  \BibitemOpen
  \bibfield  {author} {\bibinfo {author} {\bibfnamefont {D.~M.}\ \bibnamefont
  {Kaplan}}\ and\ \bibinfo {author} {\bibfnamefont {J.}~\bibnamefont
  {Michelson}},\ }\href {\doibase 10.1016/S0370-2693(97)01031-9} {\bibfield
  {journal} {\bibinfo  {journal} {Phys. Lett. B}\ }\textbf {\bibinfo {volume}
  {410}},\ \bibinfo {pages} {125} (\bibinfo {year} {1997})},\ \Eprint
  {http://arxiv.org/abs/hep-th/9707021} {arXiv:hep-th/9707021} \BibitemShut
  {NoStop}%
\bibitem [{\citenamefont {Michelson}(1998)}]{Michelson:1997vq}%
  \BibitemOpen
  \bibfield  {author} {\bibinfo {author} {\bibfnamefont {J.}~\bibnamefont
  {Michelson}},\ }\href {\doibase 10.1103/PhysRevD.57.1092} {\bibfield
  {journal} {\bibinfo  {journal} {Phys. Rev. D}\ }\textbf {\bibinfo {volume}
  {57}},\ \bibinfo {pages} {1092} (\bibinfo {year} {1998})},\ \Eprint
  {http://arxiv.org/abs/hep-th/9708091} {arXiv:hep-th/9708091} \BibitemShut
  {NoStop}%
\bibitem [{\citenamefont {DeWitt}(1967)}]{DeWitt:1967yk}%
  \BibitemOpen
  \bibfield  {author} {\bibinfo {author} {\bibfnamefont {B.~S.}\ \bibnamefont
  {DeWitt}},\ }\href {\doibase 10.1103/PhysRev.160.1113} {\bibfield  {journal}
  {\bibinfo  {journal} {Phys. Rev.}\ }\textbf {\bibinfo {volume} {160}},\
  \bibinfo {pages} {1113} (\bibinfo {year} {1967})}\BibitemShut {NoStop}%
\bibitem [{\citenamefont {Gil-Medrano}\ and\ \citenamefont
  {Michor}(1992)}]{gilmedrano1992riemannian}%
  \BibitemOpen
  \bibfield  {author} {\bibinfo {author} {\bibfnamefont {O.}~\bibnamefont
  {Gil-Medrano}}\ and\ \bibinfo {author} {\bibfnamefont {P.~W.}\ \bibnamefont
  {Michor}},\ }\href@noop {} {\enquote {\bibinfo {title} {The riemannian
  manifold of all riemannian metrics},}\ } (\bibinfo {year} {1992}),\ \Eprint
  {http://arxiv.org/abs/math/9201259} {arXiv:math/9201259 [math.DG]}
  \BibitemShut {NoStop}%
\bibitem [{\citenamefont {Lüst}\ \emph {et~al.}(2019)\citenamefont {Lüst},
  \citenamefont {Palti},\ and\ \citenamefont {Vafa}}]{Lust:2019zwm}%
  \BibitemOpen
  \bibfield  {author} {\bibinfo {author} {\bibfnamefont {D.}~\bibnamefont
  {Lüst}}, \bibinfo {author} {\bibfnamefont {E.}~\bibnamefont {Palti}}, \ and\
  \bibinfo {author} {\bibfnamefont {C.}~\bibnamefont {Vafa}},\ }\href {\doibase
  10.1016/j.physletb.2019.134867} {\bibfield  {journal} {\bibinfo  {journal}
  {Phys. Lett. B}\ }\textbf {\bibinfo {volume} {797}},\ \bibinfo {pages}
  {134867} (\bibinfo {year} {2019})},\ \Eprint
  {http://arxiv.org/abs/1906.05225} {arXiv:1906.05225 [hep-th]} \BibitemShut
  {NoStop}%
\bibitem [{\citenamefont {Candelas}\ and\ \citenamefont {de~la
  Ossa}(1991)}]{Candelas:1990pi}%
  \BibitemOpen
  \bibfield  {author} {\bibinfo {author} {\bibfnamefont {P.}~\bibnamefont
  {Candelas}}\ and\ \bibinfo {author} {\bibfnamefont {X.}~\bibnamefont {de~la
  Ossa}},\ }\href {\doibase 10.1016/0550-3213(91)90122-E} {\bibfield  {journal}
  {\bibinfo  {journal} {Nucl. Phys. B}\ }\textbf {\bibinfo {volume} {355}},\
  \bibinfo {pages} {455} (\bibinfo {year} {1991})}\BibitemShut {NoStop}%
\bibitem [{\citenamefont {Ooguri}\ and\ \citenamefont
  {Vafa}(2007)}]{Ooguri:2006in}%
  \BibitemOpen
  \bibfield  {author} {\bibinfo {author} {\bibfnamefont {H.}~\bibnamefont
  {Ooguri}}\ and\ \bibinfo {author} {\bibfnamefont {C.}~\bibnamefont {Vafa}},\
  }\href {\doibase 10.1016/j.nuclphysb.2006.10.033} {\bibfield  {journal}
  {\bibinfo  {journal} {Nucl. Phys. B}\ }\textbf {\bibinfo {volume} {766}},\
  \bibinfo {pages} {21} (\bibinfo {year} {2007})},\ \Eprint
  {http://arxiv.org/abs/hep-th/0605264} {arXiv:hep-th/0605264} \BibitemShut
  {NoStop}%
\bibitem [{\citenamefont {Klaewer}\ and\ \citenamefont
  {Palti}(2017)}]{Klaewer:2016kiy}%
  \BibitemOpen
  \bibfield  {author} {\bibinfo {author} {\bibfnamefont {D.}~\bibnamefont
  {Klaewer}}\ and\ \bibinfo {author} {\bibfnamefont {E.}~\bibnamefont
  {Palti}},\ }\href {\doibase 10.1007/JHEP01(2017)088} {\bibfield  {journal}
  {\bibinfo  {journal} {JHEP}\ }\textbf {\bibinfo {volume} {01}},\ \bibinfo
  {pages} {088} (\bibinfo {year} {2017})},\ \Eprint
  {http://arxiv.org/abs/1610.00010} {arXiv:1610.00010 [hep-th]} \BibitemShut
  {NoStop}%
\bibitem [{\citenamefont {Baume}\ and\ \citenamefont
  {Palti}(2016)}]{Baume:2016psm}%
  \BibitemOpen
  \bibfield  {author} {\bibinfo {author} {\bibfnamefont {F.}~\bibnamefont
  {Baume}}\ and\ \bibinfo {author} {\bibfnamefont {E.}~\bibnamefont {Palti}},\
  }\href {\doibase 10.1007/JHEP08(2016)043} {\bibfield  {journal} {\bibinfo
  {journal} {JHEP}\ }\textbf {\bibinfo {volume} {08}},\ \bibinfo {pages} {043}
  (\bibinfo {year} {2016})},\ \Eprint {http://arxiv.org/abs/1602.06517}
  {arXiv:1602.06517 [hep-th]} \BibitemShut {NoStop}%
\bibitem [{\citenamefont {Bena}\ and\ \citenamefont
  {Warner}(2006)}]{Bena:2005va}%
  \BibitemOpen
  \bibfield  {author} {\bibinfo {author} {\bibfnamefont {I.}~\bibnamefont
  {Bena}}\ and\ \bibinfo {author} {\bibfnamefont {N.~P.}\ \bibnamefont
  {Warner}},\ }\href {\doibase 10.1103/PhysRevD.74.066001} {\bibfield
  {journal} {\bibinfo  {journal} {Phys. Rev. D}\ }\textbf {\bibinfo {volume}
  {74}},\ \bibinfo {pages} {066001} (\bibinfo {year} {2006})},\ \Eprint
  {http://arxiv.org/abs/hep-th/0505166} {arXiv:hep-th/0505166} \BibitemShut
  {NoStop}%
\bibitem [{\citenamefont {Bena}\ \emph {et~al.}(2007)\citenamefont {Bena},
  \citenamefont {Wang},\ and\ \citenamefont {Warner}}]{Bena:2006is}%
  \BibitemOpen
  \bibfield  {author} {\bibinfo {author} {\bibfnamefont {I.}~\bibnamefont
  {Bena}}, \bibinfo {author} {\bibfnamefont {C.-W.}\ \bibnamefont {Wang}}, \
  and\ \bibinfo {author} {\bibfnamefont {N.~P.}\ \bibnamefont {Warner}},\
  }\href {\doibase 10.1103/PhysRevD.75.124026} {\bibfield  {journal} {\bibinfo
  {journal} {Phys. Rev. D}\ }\textbf {\bibinfo {volume} {75}},\ \bibinfo
  {pages} {124026} (\bibinfo {year} {2007})},\ \Eprint
  {http://arxiv.org/abs/hep-th/0604110} {arXiv:hep-th/0604110} \BibitemShut
  {NoStop}%
\bibitem [{\citenamefont {Warner}(2019)}]{Warner:2019jll}%
  \BibitemOpen
  \bibfield  {author} {\bibinfo {author} {\bibfnamefont {N.~P.}\ \bibnamefont
  {Warner}},\ }\href@noop {} {\  (\bibinfo {year} {2019})},\ \Eprint
  {http://arxiv.org/abs/1912.13108} {arXiv:1912.13108 [hep-th]} \BibitemShut
  {NoStop}%
\bibitem [{\citenamefont {Breckenridge}\ \emph {et~al.}(1997)\citenamefont
  {Breckenridge}, \citenamefont {Myers}, \citenamefont {Peet},\ and\
  \citenamefont {Vafa}}]{Breckenridge:1996is}%
  \BibitemOpen
  \bibfield  {author} {\bibinfo {author} {\bibfnamefont {J.}~\bibnamefont
  {Breckenridge}}, \bibinfo {author} {\bibfnamefont {R.~C.}\ \bibnamefont
  {Myers}}, \bibinfo {author} {\bibfnamefont {A.}~\bibnamefont {Peet}}, \ and\
  \bibinfo {author} {\bibfnamefont {C.}~\bibnamefont {Vafa}},\ }\href {\doibase
  10.1016/S0370-2693(96)01460-8} {\bibfield  {journal} {\bibinfo  {journal}
  {Phys. Lett. B}\ }\textbf {\bibinfo {volume} {391}},\ \bibinfo {pages} {93}
  (\bibinfo {year} {1997})},\ \Eprint {http://arxiv.org/abs/hep-th/9602065}
  {arXiv:hep-th/9602065} \BibitemShut {NoStop}%
\bibitem [{\citenamefont {Bena}\ \emph {et~al.}(2006)\citenamefont {Bena},
  \citenamefont {Wang},\ and\ \citenamefont {Warner}}]{Bena:2006kb}%
  \BibitemOpen
  \bibfield  {author} {\bibinfo {author} {\bibfnamefont {I.}~\bibnamefont
  {Bena}}, \bibinfo {author} {\bibfnamefont {C.-W.}\ \bibnamefont {Wang}}, \
  and\ \bibinfo {author} {\bibfnamefont {N.~P.}\ \bibnamefont {Warner}},\
  }\href {\doibase 10.1088/1126-6708/2006/11/042} {\bibfield  {journal}
  {\bibinfo  {journal} {JHEP}\ }\textbf {\bibinfo {volume} {11}},\ \bibinfo
  {pages} {042} (\bibinfo {year} {2006})},\ \Eprint
  {http://arxiv.org/abs/hep-th/0608217} {arXiv:hep-th/0608217} \BibitemShut
  {NoStop}%
\bibitem [{\citenamefont {Bena}\ and\ \citenamefont
  {Warner}(2008)}]{Bena:2007kg}%
  \BibitemOpen
  \bibfield  {author} {\bibinfo {author} {\bibfnamefont {I.}~\bibnamefont
  {Bena}}\ and\ \bibinfo {author} {\bibfnamefont {N.~P.}\ \bibnamefont
  {Warner}},\ }\href {\doibase 10.1007/978-3-540-79523-0\_1} {\bibfield
  {journal} {\bibinfo  {journal} {Lect. Notes Phys.}\ }\textbf {\bibinfo
  {volume} {755}},\ \bibinfo {pages} {1} (\bibinfo {year} {2008})},\ \Eprint
  {http://arxiv.org/abs/hep-th/0701216} {arXiv:hep-th/0701216} \BibitemShut
  {NoStop}%
\bibitem [{\citenamefont {Bates}\ and\ \citenamefont
  {Denef}(2011)}]{Bates:2003vx}%
  \BibitemOpen
  \bibfield  {author} {\bibinfo {author} {\bibfnamefont {B.}~\bibnamefont
  {Bates}}\ and\ \bibinfo {author} {\bibfnamefont {F.}~\bibnamefont {Denef}},\
  }\href {\doibase 10.1007/JHEP11(2011)127} {\bibfield  {journal} {\bibinfo
  {journal} {JHEP}\ }\textbf {\bibinfo {volume} {11}},\ \bibinfo {pages} {127}
  (\bibinfo {year} {2011})},\ \Eprint {http://arxiv.org/abs/hep-th/0304094}
  {arXiv:hep-th/0304094} \BibitemShut {NoStop}%
\bibitem [{\citenamefont {Behrndt}\ \emph {et~al.}(1998)\citenamefont
  {Behrndt}, \citenamefont {Lust},\ and\ \citenamefont
  {Sabra}}]{Behrndt:1997ny}%
  \BibitemOpen
  \bibfield  {author} {\bibinfo {author} {\bibfnamefont {K.}~\bibnamefont
  {Behrndt}}, \bibinfo {author} {\bibfnamefont {D.}~\bibnamefont {Lust}}, \
  and\ \bibinfo {author} {\bibfnamefont {W.~A.}\ \bibnamefont {Sabra}},\ }\href
  {\doibase 10.1016/S0550-3213(97)00633-0} {\bibfield  {journal} {\bibinfo
  {journal} {Nucl. Phys. B}\ }\textbf {\bibinfo {volume} {510}},\ \bibinfo
  {pages} {264} (\bibinfo {year} {1998})},\ \Eprint
  {http://arxiv.org/abs/hep-th/9705169} {arXiv:hep-th/9705169} \BibitemShut
  {NoStop}%
\bibitem [{\citenamefont {Denef}(2000)}]{Denef:2000nb}%
  \BibitemOpen
  \bibfield  {author} {\bibinfo {author} {\bibfnamefont {F.}~\bibnamefont
  {Denef}},\ }\href {\doibase 10.1088/1126-6708/2000/08/050} {\bibfield
  {journal} {\bibinfo  {journal} {JHEP}\ }\textbf {\bibinfo {volume} {08}},\
  \bibinfo {pages} {050} (\bibinfo {year} {2000})},\ \Eprint
  {http://arxiv.org/abs/hep-th/0005049} {arXiv:hep-th/0005049} \BibitemShut
  {NoStop}%
\bibitem [{\citenamefont {Bena}\ \emph {et~al.}(2018)\citenamefont {Bena},
  \citenamefont {Heidmann},\ and\ \citenamefont {Turton}}]{Bena:2018bbd}%
  \BibitemOpen
  \bibfield  {author} {\bibinfo {author} {\bibfnamefont {I.}~\bibnamefont
  {Bena}}, \bibinfo {author} {\bibfnamefont {P.}~\bibnamefont {Heidmann}}, \
  and\ \bibinfo {author} {\bibfnamefont {D.}~\bibnamefont {Turton}},\ }\href
  {\doibase 10.1007/JHEP12(2018)028} {\bibfield  {journal} {\bibinfo  {journal}
  {JHEP}\ }\textbf {\bibinfo {volume} {12}},\ \bibinfo {pages} {028} (\bibinfo
  {year} {2018})},\ \Eprint {http://arxiv.org/abs/1806.02834} {arXiv:1806.02834
  [hep-th]} \BibitemShut {NoStop}%
\bibitem [{\citenamefont {Li}(2021)}]{Li:2021gbg}%
  \BibitemOpen
  \bibfield  {author} {\bibinfo {author} {\bibfnamefont {Y.}~\bibnamefont
  {Li}},\ }\href {\doibase 10.1007/JHEP06(2021)065} {\bibfield  {journal}
  {\bibinfo  {journal} {JHEP}\ }\textbf {\bibinfo {volume} {06}},\ \bibinfo
  {pages} {065} (\bibinfo {year} {2021})},\ \Eprint
  {http://arxiv.org/abs/2102.04480} {arXiv:2102.04480 [hep-th]} \BibitemShut
  {NoStop}%
\bibitem [{\citenamefont {Denef}(2002)}]{Denef:2002ru}%
  \BibitemOpen
  \bibfield  {author} {\bibinfo {author} {\bibfnamefont {F.}~\bibnamefont
  {Denef}},\ }\href {\doibase 10.1088/1126-6708/2002/10/023} {\bibfield
  {journal} {\bibinfo  {journal} {JHEP}\ }\textbf {\bibinfo {volume} {10}},\
  \bibinfo {pages} {023} (\bibinfo {year} {2002})},\ \Eprint
  {http://arxiv.org/abs/hep-th/0206072} {arXiv:hep-th/0206072} \BibitemShut
  {NoStop}%
\bibitem [{\citenamefont {Maldacena}\ \emph {et~al.}(1999)\citenamefont
  {Maldacena}, \citenamefont {Michelson},\ and\ \citenamefont
  {Strominger}}]{Maldacena:1998uz}%
  \BibitemOpen
  \bibfield  {author} {\bibinfo {author} {\bibfnamefont {J.~M.}\ \bibnamefont
  {Maldacena}}, \bibinfo {author} {\bibfnamefont {J.}~\bibnamefont
  {Michelson}}, \ and\ \bibinfo {author} {\bibfnamefont {A.}~\bibnamefont
  {Strominger}},\ }\href {\doibase 10.1088/1126-6708/1999/02/011} {\bibfield
  {journal} {\bibinfo  {journal} {JHEP}\ }\textbf {\bibinfo {volume} {02}},\
  \bibinfo {pages} {011} (\bibinfo {year} {1999})},\ \Eprint
  {http://arxiv.org/abs/hep-th/9812073} {arXiv:hep-th/9812073} \BibitemShut
  {NoStop}%
\bibitem [{\citenamefont {Bonnefoy}\ \emph {et~al.}(2020)\citenamefont
  {Bonnefoy}, \citenamefont {Ciambelli}, \citenamefont {L\"ust},\ and\
  \citenamefont {L\"ust}}]{Bonnefoy:2019nzv}%
  \BibitemOpen
  \bibfield  {author} {\bibinfo {author} {\bibfnamefont {Q.}~\bibnamefont
  {Bonnefoy}}, \bibinfo {author} {\bibfnamefont {L.}~\bibnamefont {Ciambelli}},
  \bibinfo {author} {\bibfnamefont {D.}~\bibnamefont {L\"ust}}, \ and\ \bibinfo
  {author} {\bibfnamefont {S.}~\bibnamefont {L\"ust}},\ }\href {\doibase
  10.1016/j.nuclphysb.2020.115112} {\bibfield  {journal} {\bibinfo  {journal}
  {Nucl. Phys. B}\ }\textbf {\bibinfo {volume} {958}},\ \bibinfo {pages}
  {115112} (\bibinfo {year} {2020})},\ \Eprint
  {http://arxiv.org/abs/1912.07453} {arXiv:1912.07453 [hep-th]} \BibitemShut
  {NoStop}%
\end{thebibliography}%

\end{document}